\documentclass[12pt,preprint]{aastex}

\begin{document}
\slugcomment{Submitted to ApJ Letter}

\title{A Two Hour Quasi-Period in an Ultra-luminous X-Ray source in NGC628}

\author{Ji-Feng Liu, Joel N. Bregman, Ed Lloyd-Davies, Jimmy Irwin, Catherine
Espaillat, and Patrick Seitzer}

\affil{Astronomy Department, University of Michigan, MI 48109}

\begin{abstract}

Quasi-periodic oscillations and X-ray spectroscopy are powerful probes of black
hole masses and accretion disks, and here we apply these diagnostics to an
ultraluminous X-ray source (ULX) in the spiral galaxy NGC628 (M74).  This
object was observed four times over two years with the Chandra X-ray
Observatory and XMM-Newton, with three long observations showing dramatic
variability, distinguished by a series of outbursts with a quasi-period (QPO)
of 4,000-7,000 seconds.  This is unique behavior among both ULXs and Galactic
X-ray binaries due to the combination of its burst-like peaks and deep troughs,
its long quasi-periods, its high variation amplitudes of $>90$\%, and its
substantial variability between observations.  The X-ray spectra is fitted by
an absorbed accretion disk plus a power-law component, suggesting the ULX was
in a spectral state analogous to the Low Hard state or the Very High state of
Galactic black hole X-ray binaries.  A black hole mass of
$\sim2$--$20\times10^3 M_\odot$ is estimated from the $f_b$--$M_\bullet$
scaling relation found in the Galactic X-ray binaries and active galactic
nuclei.

\end{abstract}



\section{Introduction}

Ultra-luminous X-ray sources (ULXs), found in many nearby galaxies by EINSTEIN,
ROSAT, and recently by Chandra and XMM, are non-nuclear X-ray point sources
with X-ray luminosities in the range of $10^{39}$ -- $10^{41}$ erg/sec.
If ULXs emit at the Eddington limit, such luminosities imply black holes (BHs)
of masses 10--1000 $M_\odot$, or 100--10,000 $M_\odot$ if the sources emit at
10\% level of the Eddington luminosity.
Such intermediate mass black holes (IMBHs) can naturally explain the observed
high luminosities and, if they exist, bridge the gap between stellar mass
BHs ($\sim10^1 M_\odot$) and supermassive BHs ($10^6-10^9 M_\odot$) in the
center of galaxies.
However, such IMBHs are not predicted to be the products of ordinary stellar
evolution models, and it is still in debate whether they can form in dense
stellar fields via runaway stellar collisions with seed BHs of a few hundred $M_\odot$
(e.g., Portegies Zwart 2002).
Alternatively, these sources can be stellar mass black holes, for which the
formation is well established in theory and observation.
Special mechanisms are required for these stellar mass black holes to emit at
super-Eddington luminosities.
Such a mechanism could be extreme relativistic beaming (Georganopoulos et al.
2002), mild geometric beaming (e.g., King et al. 2001; Abramowics et al. 1978),
or the photon-bubble instability in a radiation pressure dominated accretion
disk, which leads to truly super-Eddington luminosities (Begelman 2002).

X-ray timing and spectral analyses have been widely used in the study of bright
Galactic X-ray binaries.
Combined with X-ray spectroscopy, rapid variability provides direct information
on the compact object and the accretion disk, the site of X-ray emission.
Most Galactic X-ray binaries show variability characterized as periodic
variations, quasi-periodic oscillations, and low frequency noise on times
scales from $10^{-4}$ seconds to years.  
They emit in well-defined spectral states including the Very High (VH) state,
the high/soft (HS) state, the low/hard (LH) state, and the quiescent state.
(e.g., Lewin, Van Paradijs \& Van den Heuvel 1995, and references therein).
In comparison, ULXs 
are less variable on short times scales. Swartz et al.  (2004) found in a
recent Chandra survey that only 5\%--15\% of the ULXs showed significant
variability during their Chandra exposures. In another Chandra survey, Roberts
et al. (2003) detected only short term flux variability in one of the two
epochs for two out of the eight ULXs in their survey.

Periodic variations and quasi-periodic oscillations are powerful probes of the
orbit motion and the BH -- accretion disk interaction, but have been
detected only for a few ULXs so far.
A ULX in IC342 showed small amplitude (5\%) periodic variations on a time scale
of 30--40 hours in a 155 hour ASCA observation (Sugiho et al. 2001).
A ULX in the Circinus galaxy showed X-ray eclipses and a period of 7.5 hours,
though it is possibly a foreground interloper in our galaxy given its proximity
to the Galactic plane and the similarity to a long period AM Her system (Bauer
et al. 2001).
A ULX in M51 showed $\sim$70\% variations in a smooth sinusoidal light curve
with a period of 7620$\pm$500 seconds in a 15 ks Chandra observation (Liu et
al. 2002), while in another 27 ks observation, the luminosity dropped by a
factor of 50 and there were too few counts to verify this periodicity.
From XMM and RXTE observations of a ULX in M82, Strohmayer \& Mushotzky (2003)
discovered a narrow QPO of 54.4$\pm$0.9 mHz ($\sim18$ seconds) with a r.m.s.
amplitude of 8.4\%, which suggests a geometrically thin accretion flow and
argues against substantial beaming.
Soria et al. (2004) found that the light curve of a ULX in NGC5408 showed flares of
amplitude of $\sim$30\% at a quasi-period of a few hundred seconds, which they
attributed to variability in the power-law component of the emission.
A ULX in the colliding galaxy NGC7714 showed medium amplitude ($\sim$30\%)
periodic variations on a time scale of 2 hours in a 15 ks XMM observation when
it was in a high/hard state with $L_X=6.6\times10^{40}$ erg/sec, while such a
period was not detected in an earlier XMM observation when it was in a low/soft
state with $L_X=4.4\times10^{40}$ erg/sec (Soria \& Motch 2004).

A ULX in NGC628 (M74), CXOU J013651.1+154547 was found to be extremely variable
with unique quasi-periodic variation patterns in recent Chandra and XMM
observations (Krauss et al. 2003). 
This source is located in a bubble nebula on a spiral arm of NGC628, a face-on
Sc spiral galaxy with a galactic latitude of -45$^\circ$ at a distance of 9.7
Mpc; it is most likely in NGC628 given the association with the nebula and
absence of bright foreground objects or known AGNs/QSOs in the vicinity.
In this paper, we analyze two Chandra observations and two XMM observations
available in the archive, and report its spectral and timing properties in
section 2. A discussion on the QPOs and the BH mass is presented in section 3.

\section{Data Analysis and Results}

There are four X-ray observations in the archive for the ULX, including two
Chandra ACIS observations, on 2001 June 19 (Observation ID 2057, 46.3
kiloseconds) and 2001 October 19 (ObsID 2058, 46.3 kiloseconds), and two XMM
EPIC observations, on 2002 February 2 (ObsID 0154350101, 36.9 kiloseconds) and
2003 January 7 (ObsID 0154350201, 24.9 kiloseconds).
To extract spectra and event lists for the ULX, we used CIAO 2.3 and CALDB
2.20 for the Chandra data, and SAS 6.0.0 for the XMM data.
%
%


For timing analysis, the event lists for the ULX in two Chandra observations
were extracted from the $3\sigma$ source regions enclosing 95\% of the source
photons as reported by the detection task WAVDETECT.
To estimate the background photon counts in the source region, we extracted the
events in a nearby region similar in shape but offset from any point sources.
It was found that the background counts are less than 3\% of the total counts.
In the two XMM EPIC observations, the ULX fell into the chip gap on the pn, so
only the two MOS observations are used for analysis.
The events were extracted from a source circle with a radius of 512 pixels
(i.e., $25\farcs6$) enclosing $\sim$90\% of the source photons, within which no
other point sources are detected in the two Chandra observations.
We took the photons within a nearby circle offset from all point sources as an
estimate of the background photons in the source circle; for the first (second)
XMM observation, the background counts are estimated to be 34\% (30\%) of the
total counts in the source circle.
The light curves for the four observations were binned and plotted in Figure 1.


Some distinct features of the light curves are immediately obvious in Figure 1.
The burst-like peaks and deep troughs alternate quasi-periodically in the first
three observations; the signal is too weak to see such features in the 4th
observation.
%
%
The changes in fluxes are violent on time scales of hours.
In ObsID 2057, there was a period of 5 kiloseconds without any photons
detected, preceded by a decrease in count rates, and followed by an increase
that leads to a large burst.
%
%
The average count rate is 7.5 (20.6) ACIS count/ks for the 1st (2nd) Chandra
observation, and 27.1 (15.0) MOS count/ks for the 3rd (4th) XMM observation.
The standard deviations from the average count rates  are 11.2/19.8/28.3/12.8
count/ks respectively, indicating variation amplitudes of 90\% or higher. 
%


To study the quasi-periodicity for the ULX, the power-density spectra (PDS)
were calculated with XRONOS 5.19 (Figure 2) for the first three observations.
The PDS for ObsID 2057 shows a peak indicative of a QPO around
$f\sim$2.5$\times$10$^{-4}$ Hz, which appears as a hump in an otherwise
declining power-law ($f^{-\alpha}$ with $\alpha$$\sim$1).  This PDS resembles that of
a Galactic BH binary in its Very High state or steep power-law (SPL) state as
described by McClintock \& Remillard (2003).
However, due to the large errors and low resolution at low frequencies, this
PDS also could be viewed as a flat-top spectrum at low frequencies that breaks
into a declining power-law with $\alpha$$\sim$1.5 around
$f_b\sim$2.5$\times$10$^{-4}$ Hz, thus resembling that of a BH binary in its LH
state or the {\it hard} state (McClintock \& Remillard, 2003).
The PDS for ObsID 2058 shows a significant broad peak that is more clearly
interpreted as a QPO within the frequency range of 1--4$\times10^{-4}$ Hz.
The PDS for the first XMM observation is similar to  that of a BH binary in its
LH state, with a break frequency (i.e., QPO frequency) around
$f\sim$1.5$\times$10$^{-4}$ Hz and a declining power-law with $\alpha$$\sim$1.2.
No significant features are found in the PDS for the second XMM observation,
partly due to its low count rate and short exposure.

A wavelet technique (e.g., Torrence \& Compo 1998; Hughes et al.  1998) is used
to search for the quasi-periods for the QPOs and has certain advantages over
the Fourier approach. 
The conventional Fourier transform may not work well to reveal quasi-periodic
signals at low frequencies, because the power associated with the data's window
can appear at low frequencies, and non-periodic outbursts will spread power
across the spectrum.
In comparison, the wavelet analysis studies the behavior of the signal on
different time scales as a function of time, and preserves the temporal
locality of the signal.
%
The global power spectra calculated from summing up the power spectra at all
times are shown in the period space to facilitate the search (Figure 3).
To assess the significance of a quasi-period in the global power spectrum of a
signal, its power is compared to the distribution of the power spectra of both
white and red noise random processes at that time scale (Torrence \& Compo
1998).
Three quasi-periods at $\sim4000$ seconds, $\sim6000$ seconds and $\sim7000$
seconds are found in the first three observations above the 99\% significance
level, i.e., the powers at the quasi-periods are higher than 99\% of the red
noise random processes.
The quasi-period at $\sim7000$ seconds (i.e., two hour) was present in all
three observations.

%


The spectra for the first three observations were fitted with three models, the
power-law model (i.e., wabs +powerlaw), the multi-color disk model (i.e.,
wabs +diskbb), and the disk plus power-law model (i.e., wabs +diskbb +powerlaw)
using XSPEC 11.2.0; no spectrum was fit for the 4th observation due to problems
with the Observational Data File (ODF).
The spectra were binned to a minimum of 15 photons per bin, and fitted in the
0.3--8 keV band (Table 1).
The best model for both Chandra observations is the disk plus power-law model.
Despite the flux change by a factor of $\sim$3, the two spectra are strikingly
similar, with similar model parameters (i.e., the same disk temperature of
$\sim$0.2 keV, and the same power-law photon index $\sim$1.5), and similar
fractional fluxes from the disk component ($\sim$1/4) and the power-law
component ($\sim$3/4). 
The spectral decomposition suggests the ULX might be in a spectral state
analogous to the Low Hard state of Galactic BH binaries, or the Very High state
given the presence of low frequency QPOs and the power-law photon index up to
$\sim$2.2 allowed by the fitting errors.
The power-law models are also acceptable for the two Chandra observations, with
similar best fit column density and photon index ($\sim$2.2).
%
%
Three emission lines in the two Chandra observations were detected at
$\sim2\sigma$ significance levels, and the details will be studied in a future
work.
For the combined MOS spectrum in the third observation, the disk plus power-law
model is unphysical since the power-law component increases with photon energy
($\Gamma\sim-2.5$); the power-law model is acceptable, and has the same model
parameters as in two Chandra observations within the errors.

%
%
%
%

For ObsID 2058, two separate spectra for the high and low flux states, 2058-H
and 2058-L, are extracted from the time intervals above and below 20 count/ks,
a sensible but non-unique division point.
The two spectra sample the photons in the bursts and troughs respectively, and
are fitted with the three models as the above.
Both the power-law model and the disk plus power-law model are acceptable for
these two spectra; in both models the parameters (column density, photon index,
and disk temperature) for the high flux spectrum are the same as those for the
low flux spectrum.
Such a similarity between these two spectra is expected only if the QPOs are
present in both the disk component and the power-law component of the X-ray
emission.
In accordance to this implication, the QPOs with similar variation amplitudes
are seen in the light curves of low-energy photons ($<1$ keV, dominated by the
disk component) and high-energy photons ($>1$ keV, dominated by the power-law
component).

\section{Discussion}



QPOs are seen in the X-ray observations for CXOU J013651.1+154547, one of the
four ULXs with possible QPOs.
The quasi-periodicity is directly visible in the light curves, and it is
present in observations spanning 2 years with flux changes by a factor of
three.
This is verified by the timing analysis with two different techniques, which
have revealed significant QPOs as peaks and/or breaks in the PDS.
Quasi-periods of $\sim$ 4--7 kiloseconds  are found from the observations, most
evident with a wavelet decomposition technique.
In both Chandra observations, the X-ray spectra are best fitted with a
multi-color disk plus a power-law component with similar parameters, in which
the disk components ($T_{in}\sim0.2$ keV) dominate the soft band and contribute
a quarter of the total flux, while the power-law components dominate the hard
band and contribute three quarters of the total flux.
This spectral shape, combined with the presence of low frequency QPOs, suggests
the ULX was in a spectral state analogous to the Low Hard state or the Very
High state of Galactic BH binaries.


Compared to QPOs in other ULXs and Galactic BH binaries, the QPO for CXOU
J013651.1+154547 is unique for the combination of its burst-like peaks and deep
troughs, its long quasi-periods, and its extreme variability.
The light curves of CXOU J013651.1+154547 exhibit substantial variability from
quasi-period to quasi-period.
Within a quasi-period of about two hours, the troughs are usually 10\% -- 20\%
of the average flux or lower, and the peaks are up to 5 times the average flux,
reaching $\sim5\times10^{39}$ erg/sec.
In comparison, QPOs for other ULXs and Galactic X-ray binaries usually have
smooth (sometimes sinusoidal) quasi-periodic patterns and r.m.s. amplitudes of
a few percent reaching up to 30\% only in rare occasions.
The QPOs are present in both the accretion disk component and the power-law
component.
As a comparison, the QPOs for an ULX in NGC 5408 with a quasi-period of a few
hundred seconds are significant only in the hard power-law component (Soria et
al.  2004).


%


The BH mass can be estimated with the $f_b$--$M_\bullet$ scaling relation if
the QPOs originate from the accretion disk as for Galactic X-ray binaries.
The scaling relation between the QPO frequency and the BH mass has been found
to exist in Galactic BH X-ray binaries and active galactic nuclei, albeit with
large uncertainties in the calibration (e.g., Belloni \& Hasinger 1990;
Markowitz et al.  2003).
The PDS for CXOU J013651.1+154547 is comparable to that of Cygnus X-1 ($\sim10
M_\odot$) which at low frequencies exhibits a flat top that breaks at $f_{b1}$
of 0.04--0.4 Hz into a declining power-law with slope $\sim-1$ that breaks
again at $f_{b2}$ of a few Hz into a steeper power-law with slope $\sim-2$.
If the observed QPO frequency $\sim2\times10^{-4}$ Hz corresponds to $f_{b1}$
as suggested by the PDS shape, the BH mass is estimated as $\sim$2--
20$\times10^3 M_\odot$.
The BH mass would be $\sim10^5 M_\odot$ if the QPO frequency corresponds to
$f_{b2}$, as is possible owing to the uncertainties in the PDS of these short
observations.
The above mass estimates are not conclusive, since the low frequency QPOs occur
at different locations of the accretion disk for different accretion rates,
thus their frequencies are not determined only by the BH mass.
This mass estimate is different from the mass estimated from its luminosities
($\le100 M_\odot$) if the system emits isotropically at above 10\% of the
Eddington limit  as indicated by the emission state possibly analogous to the
Very High state in Galactic BH binaries.


More convincing and stringent mass estimates can be made with high frequency
QPOs, especially the 3:2 twin-peak QPOs that appear as kHz QPOs in Galactic
X-ray binaries.
This 3:2 twin-peak QPO is thought to occur at a specific resonance radius in the inner
region of the accretion disk that is fixed in terms of the gravitational radius
of the  BH. Thus its frequency is expected to scale with the BH mass.
Abramowics et al. (2004) advanced a stringent scaling relation
$f=2.8/M_\bullet$ kHz for such QPOs, which is verified in three microquasars
and the central BH of our Galaxy, Sgr A*.
If the 3:2 twin-peak QPO is to occur in NGC628-ULX, we expect a frequency of
$\sim0.3$ Hz given the BH mass of $\sim10^{4} M_\odot$ as inferred from the
observed low frequency QPOs.
However, to verify such a QPO would require a lengthy observation and the
object would need to be brighter than usual.

\acknowledgements

We are grateful for the service of Chandra and XMM data archives. We would like
to thank Philip Hughes, Renato Dupke, and Eric Miller for helpful discussions.
We gratefully acknowledge support for this work from NASA under grants
HST-GO-09073.




\begin{deluxetable}{llllllllll}
\tablecaption{Spectral fits for CXOU J013651.1+154547}
\tabletypesize{\tiny}
\tablehead{
\colhead{spectrum} & \colhead{model} & \colhead{$N_H$} & \colhead{$\Gamma$} & 
\colhead{$kT_{in}$} & \colhead{$\chi^2_\nu$/dof} & \colhead{Prob} & \colhead{$L_X$(0.3-8)} & 
\colhead{$L_{diskbb}$(0.3-8)} & \colhead{$L_{pl}$(0.3-8)} \\
\colhead{}& \colhead{}& \colhead{($10^{20}$ $cm^{-2}$)} & \colhead{} & \colhead{(Kev)} &
\colhead{} & \colhead{\%} &  \colhead{($10^{38}$erg/s)} & \colhead{($10^{38}$erg/s)} & 
\colhead{($10^{38}$erg/s)}
}

\startdata

    & wabs(pl) & 6.4$\pm$4.6 & 2.4$\pm$0.3 & \nodata & 0.87/50 & 73.4 & 3.8 & \nodata & 3.8 \\
acis2057 & wabs(diskbb) & 0.0$\pm$100 & \nodata & 0.46$\pm$0.03 & 1.12/50 & 26.5 & 2.8 & 2.8 & \nodata \\
    & wabs(pl+diskbb) & 8.8$\pm$10.4 & 1.5$\pm$0.7 & 0.21$\pm$0.08 & 0.81/48 & 82.7 & 4.5 & 1.3 & 3.2 \\

\hline

   & wabs(pl) & 6.2$\pm$2.2 & 2.2$\pm$0.1 & \nodata & 1.10/109 & 22.3 & 11.6 & \nodata & 11.6 \\
acis2058 & wabs(diskbb) & 0$\pm$100 & \nodata & 0.64$\pm$0.03 & 1.53/109 & 0 & 8.4 & 8.4 & \nodata \\
   & wabs(pl+diskbb) & 7.2$\pm$4.6 & 1.4$\pm$0.4 & 0.24$\pm$0.05 & 1.04/107 & 36.4 & 13.4 & 3.3 & 10.1 \\

\hline

   & wabs(pl) & 15.8$\pm$3.2 & 2.1$\pm$0.2 & \nodata & 0.92/86 & 67.6 & 21.4 & \nodata & 21.4 \\
2058-H & wabs(diskbb) & 2.9$\pm$1.7 & \nodata & 0.93$\pm$0.09 & 1.11/86 & 22.7 & 17.8 & 17.8 & \nodata \\
   & wabs(diskbb+pl) & 28.7$\pm$13.8 & 1.8$\pm$0.3 & 0.16$\pm$0.03 & 0.88/84 & 78 & 23.4 & 3.3 & 20.1 \\

\hline

   & wabs(pl) & 9.6$\pm$4.4 & 2.3$\pm$0.3 & \nodata & 1.21/43 & 16.7 & 3.8 & \nodata & 3.8 \\
2058-L & wabs(diskbb) & 0$\pm$5.3 & \nodata & 0.74$\pm$0.10 & 1.44/43 & 3.0 & 3.2 & 3.2 & \nodata \\
   & wabs(pl+diskbb) & 9.1$\pm$9.1 & 1.7$\pm$0.9 & 0.25$\pm$0.13 & 1.22/41 & 15.1 & 4.2 & 0.9 & 3.2 \\

\hline

  & wabs(pl) & 14.3$\pm$8.9 & 2.1$\pm$0.4 & \nodata & 1.01/35 & 45.7 & 4.6 & \nodata & 4.6 \\
0154350101 & wabs(diskbb) & 0$\pm$11.3 & \nodata & 0.91$\pm$0.17 & 1.00/35 & 46.7 & 3.9 & 3.9 & \nodata \\
  & wabs(diskbb+pl) & 0$\pm$100 & -2.5$\pm$100 & 0.86$\pm$0.39 & 1.06/33 & 38.1 & 5.3 & 3.8 & 1.5 \\

\enddata

\end{deluxetable}

\begin{figure}
\plotone{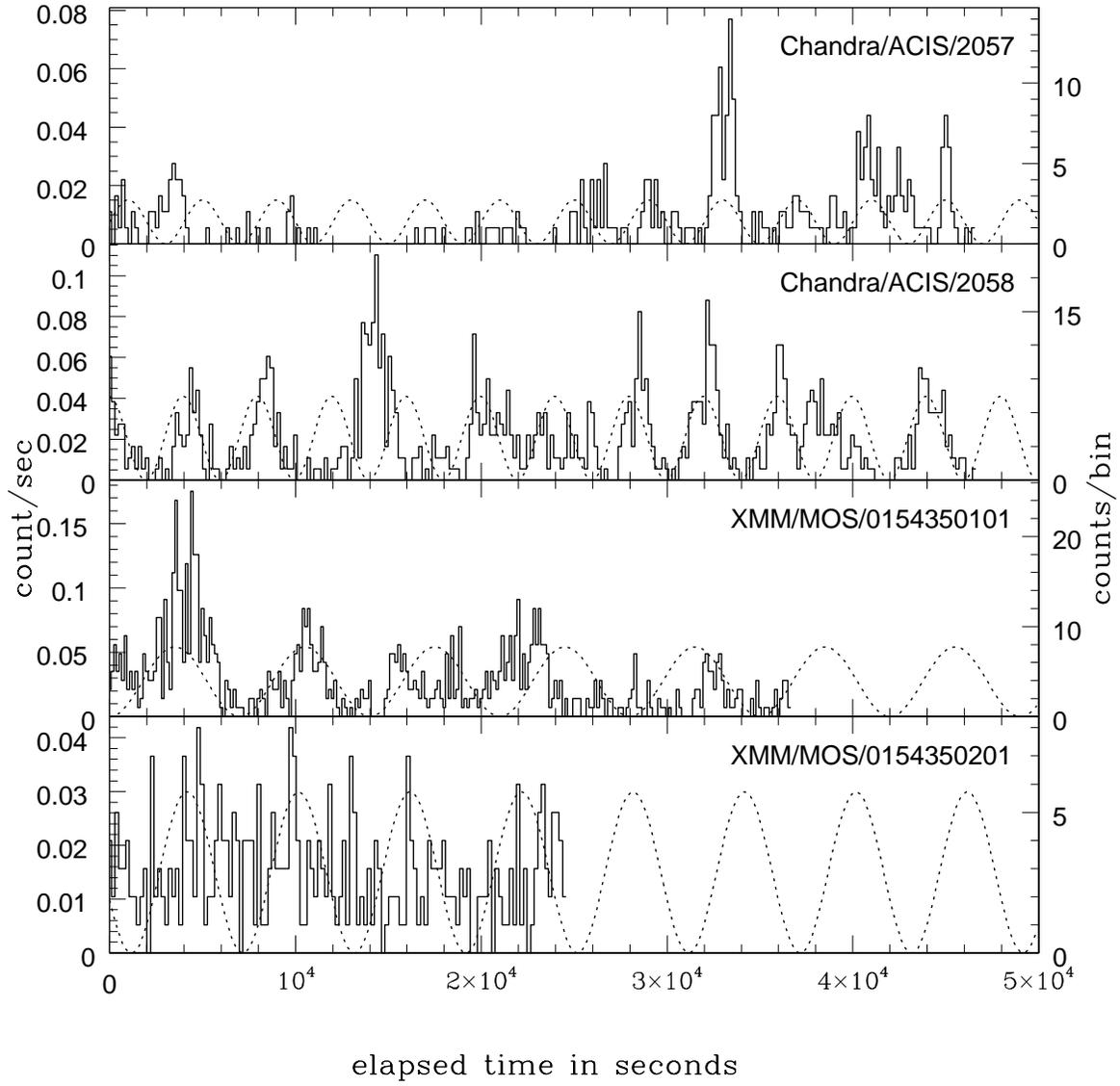}

\caption{The light curves for CXOU J013651.1+154547 in four observations. 
Overplotted for comparison are sinusoidal curves with the periods of
4000/4000/7000/6000 seconds respectively.}

\end{figure}

\begin{figure}
\plotone{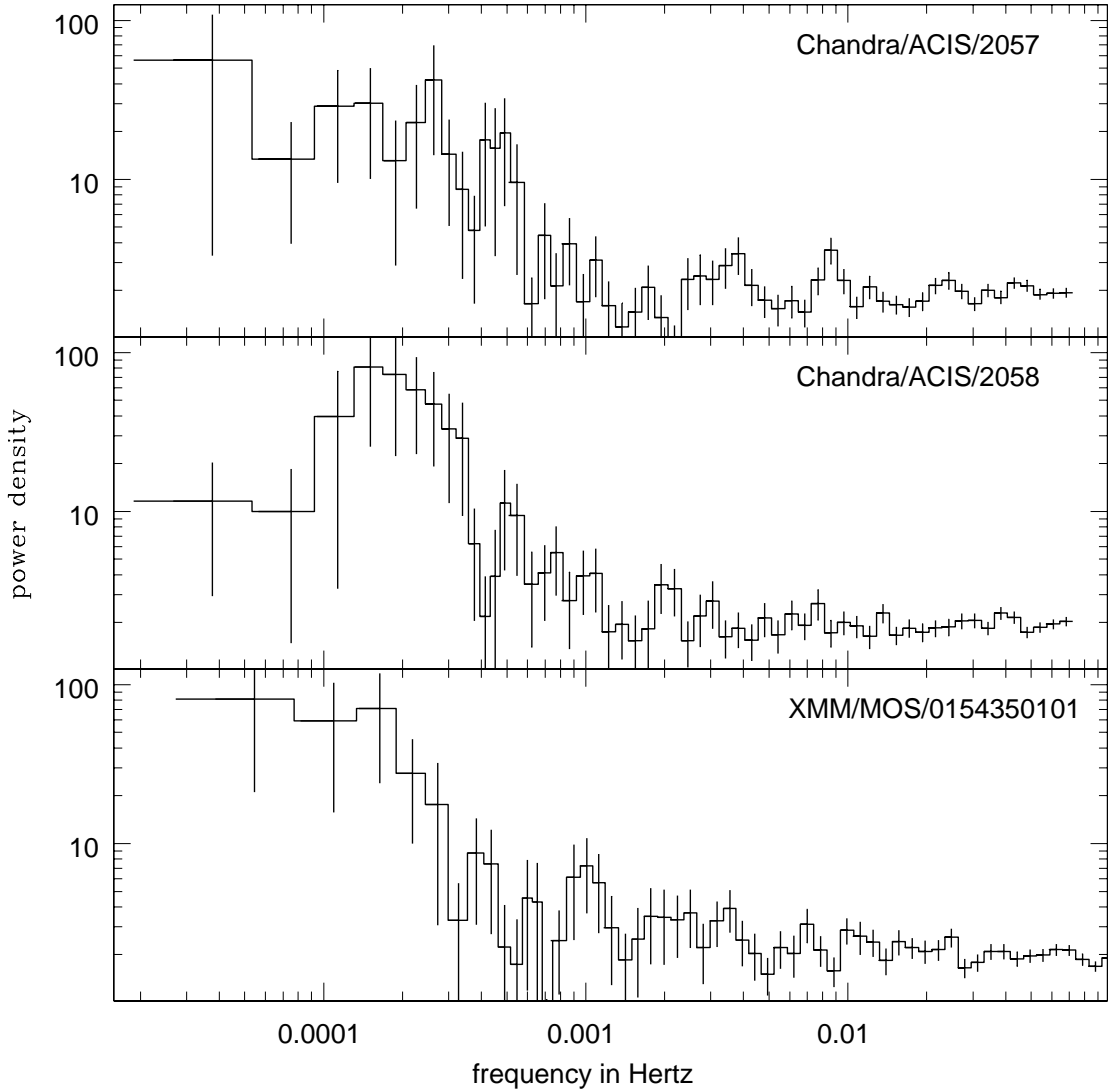}

\caption{The power-density spectra calculated with the Fourier transform for
CXOU J013651.1+154547 in the first three observations.  A broad peak around
1--4$\times10^{-4}$ Hz is clearly present in the second Chandra observation
that can be interpreted as a QPO.  }

\end{figure}

\begin{figure}
\plotone{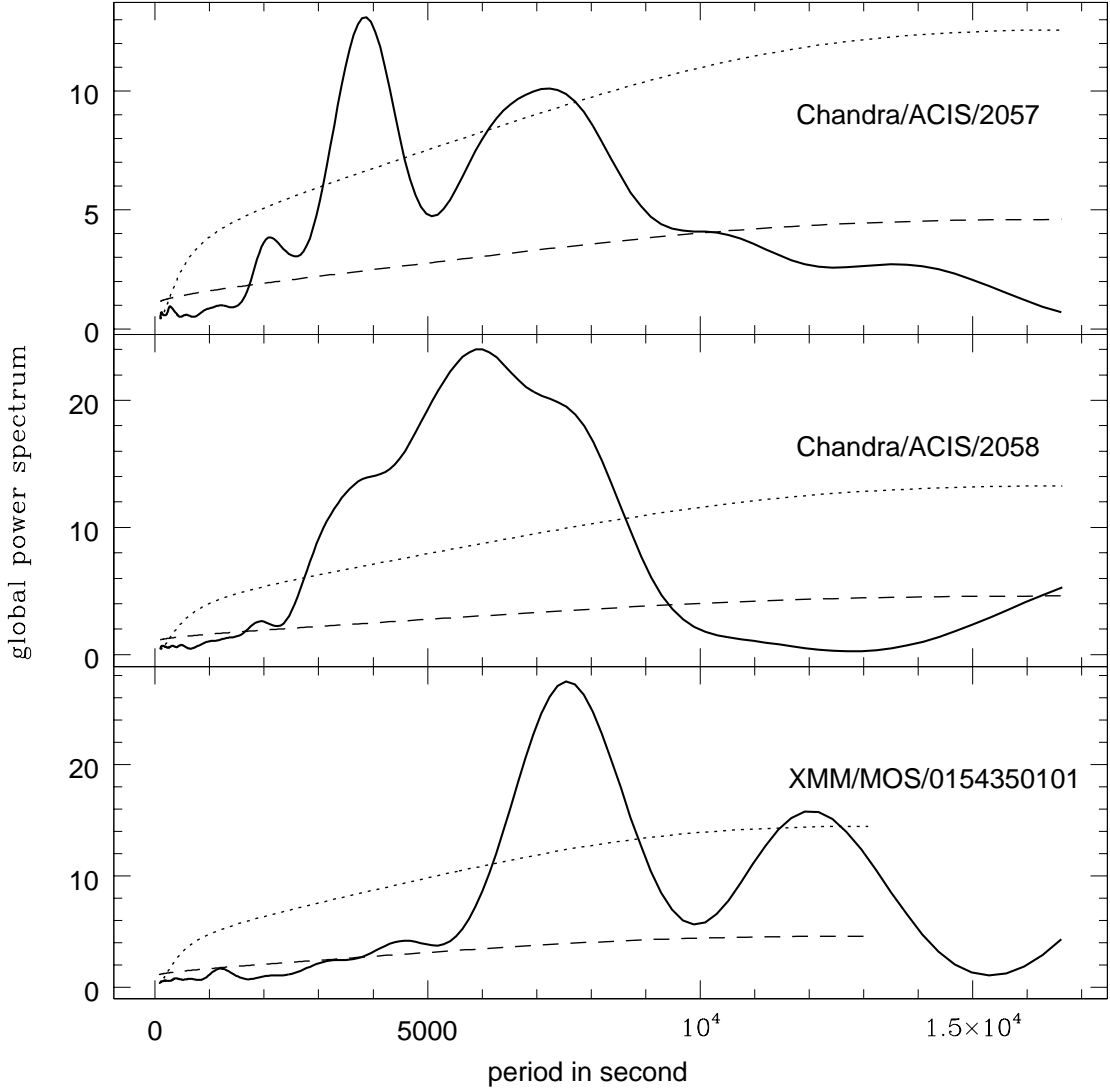}

\caption{The global power spectra calculated with a wavelet technique for CXOU
J013651.1+154547.  The dotted (dashed) line indicates the significance level at
which the power of a (quasi-)periodic signal is above 99\% of the random
realizations of a red (white) noise process.  Above the 99\% significance level
of a red noise process are quasi-periods at $\sim$4000 seconds and $\sim$7000
seconds in the first observation, quasi-periods at $\sim$4000 seconds,
$\sim$6000 seconds and $\sim$7000 seconds blended together in the second
observation, and quasi-periods at $\sim$7000 seconds and $\sim$12000 seconds in
the third observation.  The 7 kilosecond quasi-period can be seen in all three
observations.  }

\end{figure}

\end{document}